\documentclass[aps,prl,longbibliography,twocolumn,superscriptaddress,showpacs]{revtex4-1}
\usepackage{amsmath,amssymb,graphicx}

\usepackage[utf8]{inputenc}
\usepackage[T1]{fontenc}
\usepackage{xcolor}

\IfFileExists{newtxtext.sty}
   {\usepackage{newtxtext,newtxmath}}
   {\IfFileExists{stix.sty}
      {\usepackage{stix}}
      {\IfFileExists{mathptmx.sty}
      {\usepackage{mathptmx}}{} } }

\usepackage{textcomp}

\usepackage{bm}

\IfFileExists{siunitx.sty}{\usepackage{booktabs,siunitx}}{}

\pdfoutput=1
\usepackage{color}
\definecolor{LinkColor}{rgb}{0.256,0.439,0.588}
\usepackage{hyperref}
\hypersetup{
   pdfauthor={good guys},
   pdftitle={good title},
   colorlinks=true,
   citecolor=LinkColor,
   linkcolor=LinkColor,
   urlcolor=LinkColor
}

\renewcommand{\vec}[1]{\mathbf{#1}}

\usepackage{pifont}

\newcommand{\etal}{\textit{et al}.}

\usepackage{multirow}

\begin{document}

\title{Pair Density Wave in the Doped $t$-$J$ Model with Ring Exchange on a Triangular Lattice}

\author{Xiao Yan Xu}
\email{wanderxu@gmail.com}
\affiliation{Department of Physics, Hong Kong University of Science and Technology, Clear Water Bay, Hong Kong, China}
\author{K. T. Law}
\email{phlaw@ust.hk}
\affiliation{Department of Physics, Hong Kong University of Science and Technology, Clear Water Bay, Hong Kong, China}
\author{Patrick A. Lee}
\email{palee@mit.edu}
\affiliation{Department of Physics, Massachusetts Institute of Technology, Cambridge Massachusetts 02139, USA}

\date{Oct 18, 2018}

\begin{abstract}
In our previous work [Phys. Rev. Lett. \textbf{121}, 046401 (2018)], we found a quantum spin liquid phase with a spinon Fermi surface in the two dimensional spin-1/2 Heisenberg model with four-spin ring exchange  on a triangular lattice. In this work we dope the spinon Fermi surface phase by studying the $t$-$J$ model with four-spin ring exchange. We perform density matrix renormalization group calculations on four-leg cylinders of a triangular lattice and find that the dominant  pair correlation function is that of a pair density wave; i.e., it is oscillatory while decaying  with distance with a power law. The doping dependence of the period is studied. This is the first example where pair density wave is the dominant pairing in a generic strongly interacting system where the pair density wave cannot be explained as a composite order and no special symmetry is required. 
\end{abstract}

\maketitle
A pair density wave (PDW) is a superconducting state in which Cooper pairs have finite momentum. The first example of PDW is the Fulde-Ferrell-Larkin-Ovchinnikov (FFLO) state~\cite{fulde1964superconductivity,larkin1965inhomogeneous} which can arise in superconductors in strong magnetic fields when the Fermi surface is split by Zeeman effect. Recently PDW has come into prominence in the context of underdoped Cuprate superconductors.  The striped PDW was proposed as a mechanism for dynamically inter-layer decoupling observed in 1/8 hole doped La$_{2-x}$Ba$_x$CuO$_4$~\cite{himeda2002stripe,li2007two,berg2007dynamical,berg2009striped}. One of us has proposed fluctuating bi-directional PDW as the "mother state" that is responsible for many of the anomalous properties of the pseudo-gap regime~\cite{lee2014amperean}. Experimentally a direct observation of PDW has been made via local  Cooper pair tunnelling in Bi$_2$Sr$_2$CaCu$_2$O$_{8+x}$~\cite{hamidian2016detection}. They found Cooper pair density modulation with period 4$a_0$ where $a_0$ is the length of unit cell, and the magnitude of the modulation is five percent of the uniform pairing background. This may be interpreted as a subsidiary PDW being generated by the period 4$a_0$ charge order together with the uniform d-wave pairing. In this sense the recent report ~\cite{edkins2018magnetic} of short range period 8$a_0$ charge order in the vicinity of the vortex core is even more exciting, because it may be the signature of a hidden period 8$a_0$ PDW~\cite{wang2018pair,dai2018pair}. 

Theoretically, there are very few microscopic models which are shown to have PDW ground states. Berg \etal ~\cite{berg2010pair} studied a Kondo-Heisenberg model with 1D electron gas coupled to a spin chain. They found a spin gapped phase with PDW correlations oscillating with period $2a_0$, which matches the period of the ordering tendency of the spin chain.  An extended two-leg Hubbard-Heisenberg model is also found to have a spin gapped phase with a  PDW ~\cite{jaefari2012pair}. In all these examples, the PDW is commensurate and can either be interpreted as a composite order between short range spin order with the same commensurate period and another short range triplet pairing order, or requires the specially tuned symmetry of  $\pi$ flux through the ladder plaquette.  Dodaro \etal ~\cite{dodaro2017intertwined} searched for PDW in a more standard $t$-$J$ model for doped cuprates, but they did not find any evidence of  PDW ordering even when they include next nearest neighbor (NNN) hopping and NNN exchange coupling.  On a more speculative level, another mechanism to generate PDW is the Amperean pairing~\cite{Lee2007amperean,lee2014amperean}, which was first proposed for quantum spin liquids with spinon Fermi surface.  The Ampere effect of the gauge magnetic field produces attractive interactions between spinons moving in the  same directions, which creates PDW with momentum $2k_F$ at a given point on the Fermi surface. In the slave boson theory, the electron operator $c_\sigma$ is written as $b^\dagger f_\sigma$ where $f_\sigma$ represents the spinon. Upon doping, the boson $b$ acquires an expectation value in  mean field theory, and  spinon pairing immediately leads to electron pairing, in this case at finite momentum. This line of reasoning is partly what motivated us to search for PDW in the context of a doped spin liquid.

Recently, we found the spinon Fermi surface phase in
 a two dimensional spin-1/2 Heisenberg model with four-spin ring exchange on a triangular lattice~\cite{he2018spinon} using density matrix renormalization group (DMRG) method. The model  is introduced as a microscopic model for Mott insulator phase of 1T-TaS$_2$. The spinon Fermi surface phase was also proposed in early works for organic compounds~\cite{Motrunich2005, Lee2005} and confirmed in the two-leg and four-leg ladder DMRG simulations\cite{Sheng2009,Block2011}. We noticed a curious absence of spin structure factor peak along the $\Gamma$ to $M$ direction and speculated on the possibility of Amperean pairing between the spinons~\cite{he2018spinon}. It is then natural to extend this work to the doped case to see if any evidence of superconductivity emerges.  Based on DMRG calculations on four-leg ladders, we find that upon doping the spinon Fermi surface state, the dominant pairing channel has oscillatory correlations, with a period which depends smoothly on doping and therefore appears to be incommensurate. Also generically  the period does not match that of any other charge or spin order, implying that there is no simple interpretation of the PDW as a composite order.  The PDW phase we found is very unique and to the best of our knowledge it is the \textit{first} example of PDW found in a generic interaction driven one band model. This is the key finding in this work. 

\textit{Model and method} \,---\,
We consider a $t$-$J$ model with four-spin ring exchange terms on a triangular lattice, $H= \hat{\mathcal{P}} \left( H_{t-J} + H_K \right) \hat{\mathcal{P}}$, where $\hat{\mathcal{P}}$ excludes doubly occupied states. The hopping and two-spin exchange term $H_{t-J}$ and four-spin ring exchange term $H_K$ are written as
\begin{align}
& H_{t-J} =-t\sum_{\langle i,j\rangle\sigma} \left( c_{i\sigma}^{\dagger}c_{j\sigma}+\text{h.c.} \right) + J\sum_{\langle i,j\rangle}\vec{S}_{i}\cdot\vec{S}_{j}, \\
& H_{K} =K\sum_{\langle i,j,k,l\rangle}\left[\left(\vec{S}_{i}\cdot\vec{S}_{j}\right)\left(\vec{S}_{k}\cdot\vec{S}_{l}\right)+\left(\vec{S}_{j}\cdot\vec{S}_{k}\right)\left(\vec{S}_{i}\cdot\vec{S}_{l}\right)\right. \nonumber \\
 & \hskip 6em \left.-\left(\vec{S}_{i}\cdot\vec{S}_{k}\right)\left(\vec{S}_{j}\cdot\vec{S}_{l}\right)\right],
\end{align}
where $\langle i,j \rangle$ denotes nearest neighbor bond, and $\langle i,j,k,l \rangle$ runs over all compact rhombuses. The ring exchange terms simulate the proximity of the undoped insulator to the Mott transition.  We already know from our earlier work on six-leg and eight-leg ladders DMRG simulations that  the undoped system enters the spinon Fermi surface phase for $K/J>0.3$. Here we investigate the effect of hole doping.

\begin{figure}[t!]
\includegraphics[width=1.0\columnwidth]{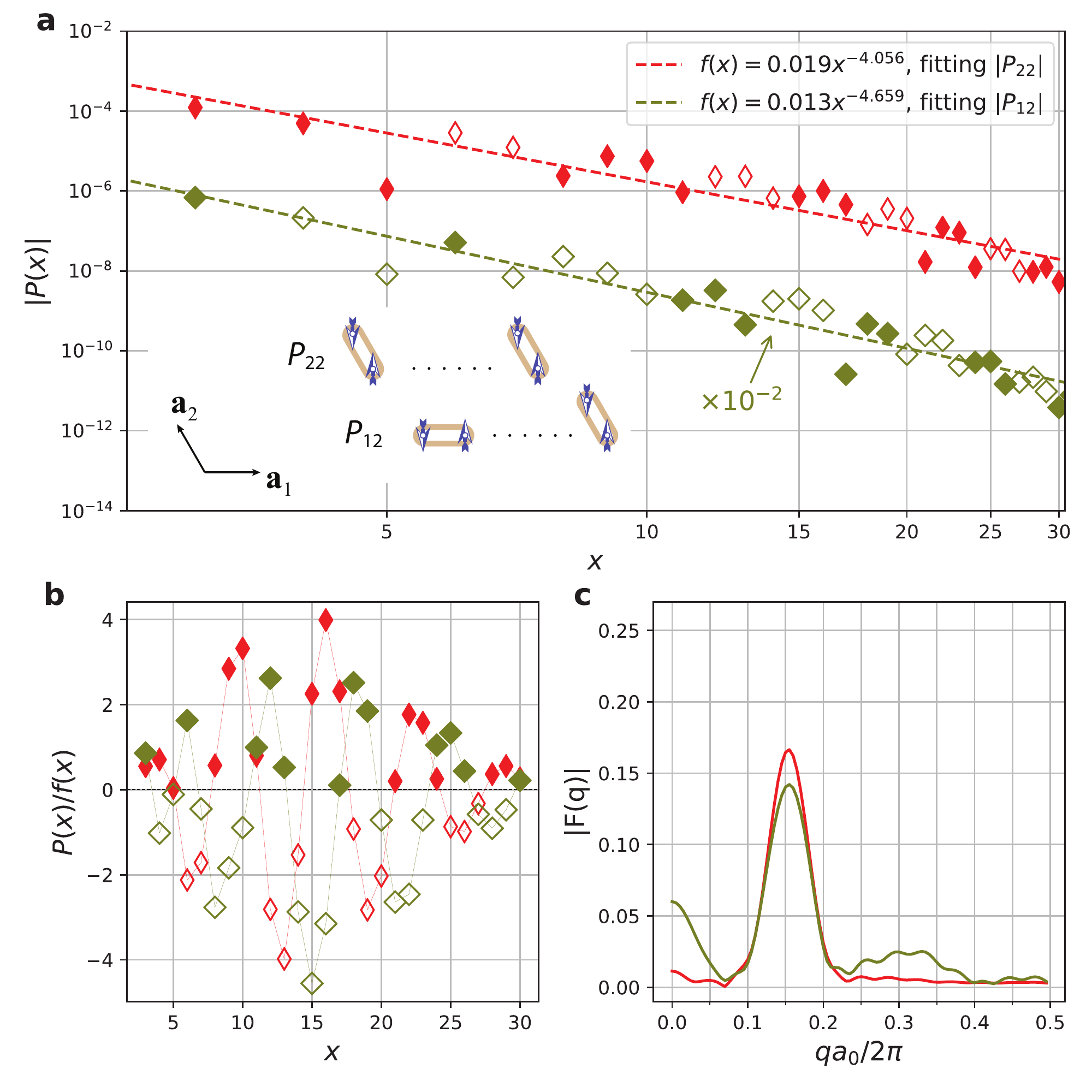}
\caption{ Pairing correlation along the long direction of the ladder for 1/16 hole doping with $K/J=0.8$, $t/J=2$ and four-leg ladder length $L_x=72$. (a) Log-log plot of the pairing correlation $P_{22}$ and $P_{12}$. Here $P_{22}$ is the correlation between pairing order parameters defined both in $\vec{a}_2$ orientation, $P_{12}$ is the correlation between pairing order parameters defined in $\vec{a}_1$ and $\vec{a}_2$ orientation, as showed in the inset. The red thin diamond is for $P_{22}$, olive diamond is for $P_{12}$, and the data for $P_{12}$ have been shifted vertically for clarity. The solid symbol is for positive value, while the open symbol is for negative value and we only plot the magnitude. The power law function $f(x)$ in the plot is a fit through the magnitude of the data points. (b) Pairing correlation $P_{22}$ and $P_{12}$ normalized by the power law function $f(x)$, which directly reflects the oscillation of the pairing correlation. The out of phase oscillation of $P_{22}$ and $P_{12}$ indicates a $d$-wave type pairing.
(c) Fourier transform of $P_{22}(x)/f(x)$ and $P_{12}(x)/f(x)$. The  peak lies at $0.15(2\pi/a_0)$ both for $P_{22}$ and $P_{12}$, which gives the total momentum of the pairing.}
\label{fig1}
\end{figure}

Soon after the discovery of high $T_c$ Cuprates, Anderson \cite{Anderson1987} proposed that doping a Mott insulator may lead to a correlation driven superconductor.  Since that time, many methods including  mean field theory, Gutzwiller variational methods and DMRG simulations and exact diagonalization have found $d$-wave type superconductivity both on square and triangular lattice in some parameter region and doping level~\cite{kotliar1988superexchange,white1997ground,wang2004doped,jiang2018,zheng2017stripe}. 
We confirm that for the standard $t$-$J$ model ($K=0$) , doping of the N\'eel ordered state produces uniform d-wave pair correlations (see Supplemental Material~\cite{suppl}). However, the situation changes completely when we dope into the spinon Fermi surface state.  We choose $K/J=0.8$ which put us quite deep into the spinon Fermi surface phase and  $t/J=2$ which is a conventional value in Mott insulator materials. We perform large-scale DMRG calculations on four-leg ladders with long direction length $L_x$ up to 72. We take periodic boundary conditions in short direction of the ladder, and open boundary conditions in the long direction. The good quantum numbers of total spin $S^{z}_{\text{tot}}$ and total number of fermions $N_{\text{tot}}$ are used. All the calculations are performed in the $S^{z}_{\text{tot}}=0$ and  $N_{\text{tot}}=N(1-p)$ sector, where $N$ is the total number of sites and $p$ is the doping level. The calculations are performed with bond dimensions up to $m=5120$ and corresponding truncation error is less than $10^{-5}$. All the results shown in the following are after extrapolating to infinite bond dimension if it is not specified. More details are presented in the Supplemental Material~\cite{suppl}.

\begin{figure}[t!]
\includegraphics[width=1.0\columnwidth]{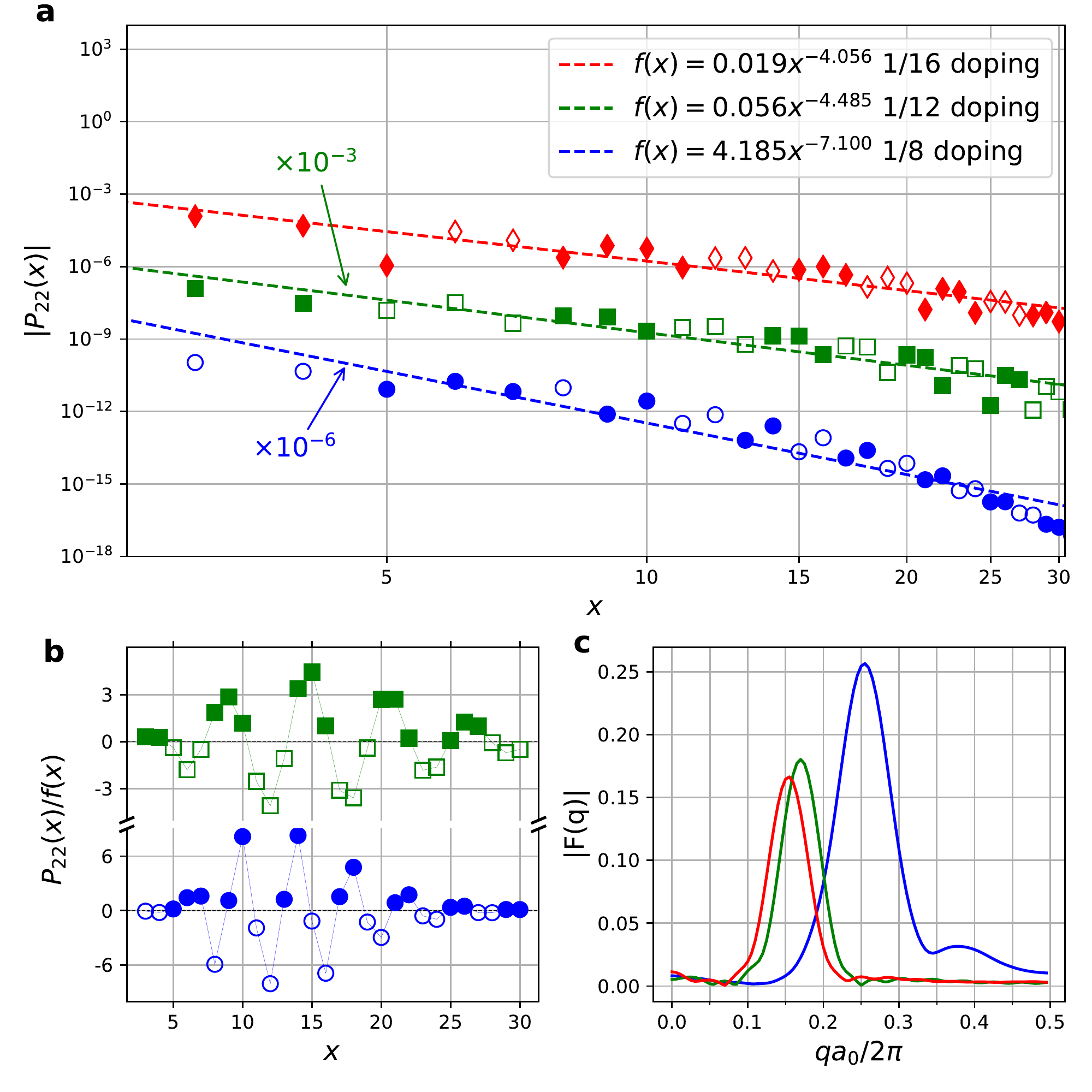}
\caption{Pairing correlation $P_{22}$ for different hole doping, 1/16, 1/12 and 1/8, with parameters $K/J=0.8$, $t/J=2$ and four-leg ladder length $L_x=72$. (a) Power law fitting of the magnitude of pairing correlation $P_{22}$ for each doping. The data for 1/12 and 1/8 hole doping have been shifted vertically for clarity. (b) Normalized pairing correlation which shows the oscillation part of $P_{22}$ for doping 1/12 and 1/8, while 1/16 doping is already showed in Fig.~\ref{fig1}(b). (c) Fourier transformation of the oscillation part of $P_{22}$ for each doping, which gives pairing momentum about $0.15(2\pi/a_0)$, $0.17(2\pi/a_0)$ and $0.25(2\pi/a_0)$ for  1/16, 1/12 and 1/8 hole doping, respectively.
}
\label{fig2}
\end{figure}

\begin{figure}[t!]
\includegraphics[width=1.0\columnwidth]{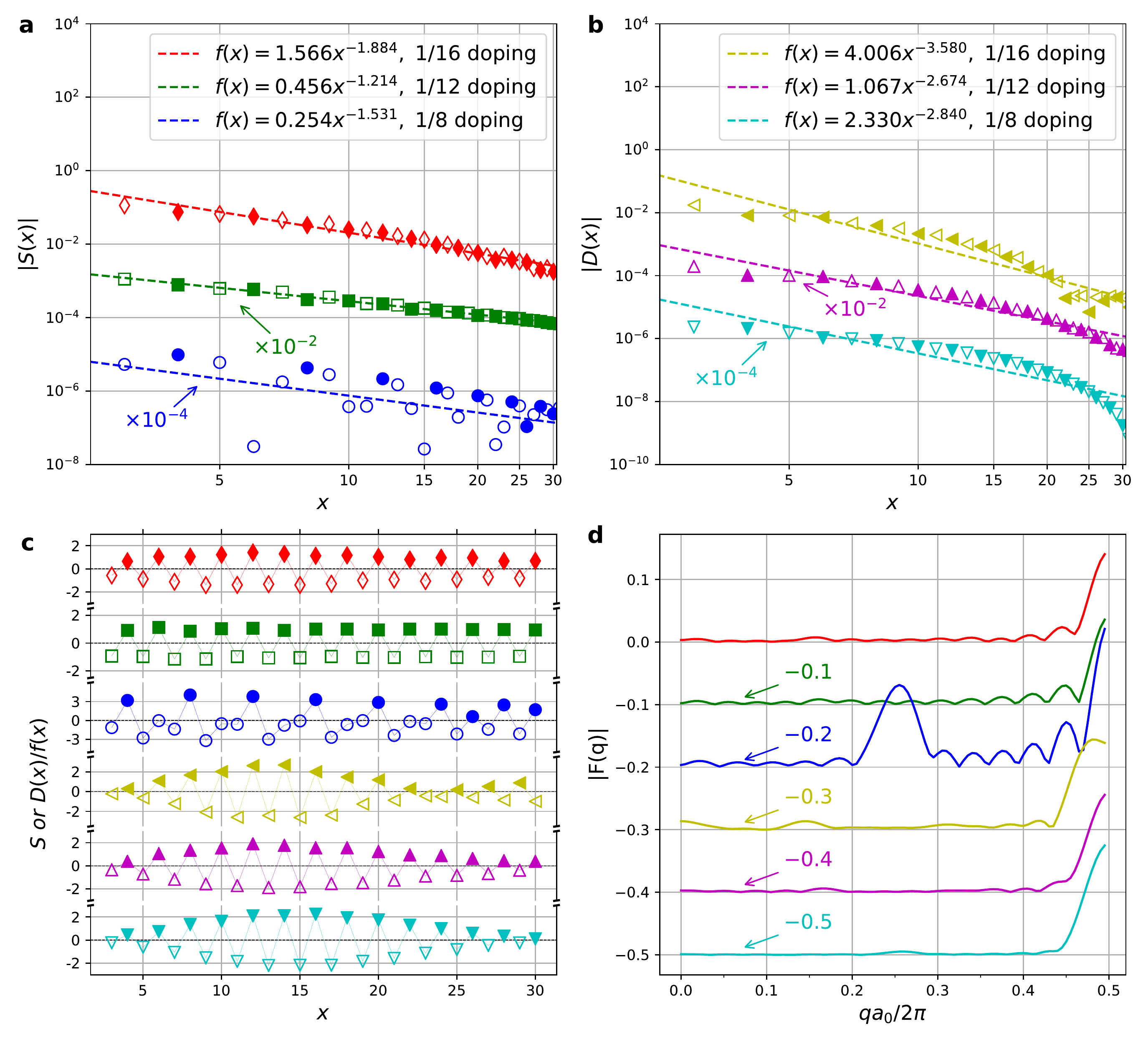}
\caption{Spin correlation $S$ and dimer correlation $D$ for different hole doping, 1/16, 1/12 and 1/8, with parameters $K/J=0.8$, $t/J=2$ and four-leg ladder length $L_x=72$.
(a) Spin correlation $S$. Red thin diamond, green square and blue circle are for 1/16, 1/12 and 1/8 hole doping respectively. Solid symbol is for positive value, open symbol is for negative value. The magnitude is plotted, and the data points are fitted with a power law. Note the vertical shift of the data for different doping for clarity.
(b) Same as (a) for dimer correlation $D$. Yellow left triangular, magenta up and cyan down are for 1/16, 1/12 and 1/8 hole doping respectively. Solid symbol is for positive value, open symbol is for negative value.
(c) Normalization with the power law function of the magnitude, giving the oscillation part of spin and dimer correlations. 
(d) Fourier transformation of the the oscillatory part  of the spin and dimer correlations. The peak gives the ordering momentum, it is $0.25(2\pi/a_0)$ for the spin correlation and $0.5(2\pi/a_0)$ for the dimer correlation at 1/8 hole doping. For 1/12 and 1/16 hole doping it is  $0.5(2\pi/a_0)$ for both correlations.
}
\label{fig3}
\end{figure}
\textit{Results}\,---\,
To study the pairing properties, we measured the pairing correlation in real space. We considered both spin singlet and triplet pairing, and find the magnitude of singlet pairing is always larger than the triplet one, so in the following we will focus on singlet pairing data. The singlet pairing order parameter is defined on nearest neighbor bonds $\Delta_{a}(i)=c_{i,\uparrow} c_{i+\delta_{a}, \downarrow} - c_{i,\downarrow} c_{i+\delta_{a}, \uparrow}$, where $\delta_{a}$ with $a=1,2$ takes the value of primitive vectors $\vec{a}_1$ and $\vec{a}_2$ respectively, denoting different orientation of pairs as shown in the inset of Fig.~\ref{fig1}(a).
To reduce finite size effects, we measure the correlation functions with a summation over the short direction $P_{aa'}(i_x-i_{x_{0}})=\sum_{i_y}\left\langle\Delta_{a}^{\dagger}(i_{x_{0}},i_y)\Delta_{a'}(i_x,i_y)\right\rangle$, where $(i_x,i_y)$ is the coordinate of site $i$ in the unit of primitive vectors, $i_{x_0}$ is a reference coordinate in the long direction, we take $i_{x_0}>\frac{L_x}{4}$ to reduce boundary effect, and the final results are averaged over several $i_{x_0}$s. 
We denote the relative distance in the correlator by $x=i_x-i_{x_{0}}$. Fig.~\ref{fig1} shows the pairing correlation at doping 1/16 for both $P_{22}$ and $P_{12}$. $P_{11}$ is also calculated, but it is very small and is not shown. The  pairing correlation  shows a clear oscillation with an amplitude which is consistent with a power law decay over a large range of $x$. (We attribute the deviation from a power law for large $x$ to finite size effect and a lack of convergence.) In the figure, the fit is made to the amplitude of the individual data points which tend to overestimate the exponent of the power law decay. For an oscillatory function, the proper way to fit the exponent requires first extracting the envelop function which we have not attempted here. The purpose of the fit we did is to allow us to display the oscillations on a linear scale, as is done in Fig.~\ref{fig1}(b). While the value of the exponent from the fit should not be taken seriously, it is apparently larger than 2, which means that the Fourier transform of the response function will not show divergence for small $q$ and $ \omega $.
For  larger doping of 1/12 and 1/8, we also see similar oscillation behavior, but with faster decay and shorter period with increasing doping (see Fig.~\ref{fig2}).  The  periods are about  $7a_0$, $6a_0$, and $4a_0$ for 1/16, 1/12, 1/8 doping respectively.

\begin{figure}[t!]
\includegraphics[width=1.0\columnwidth]{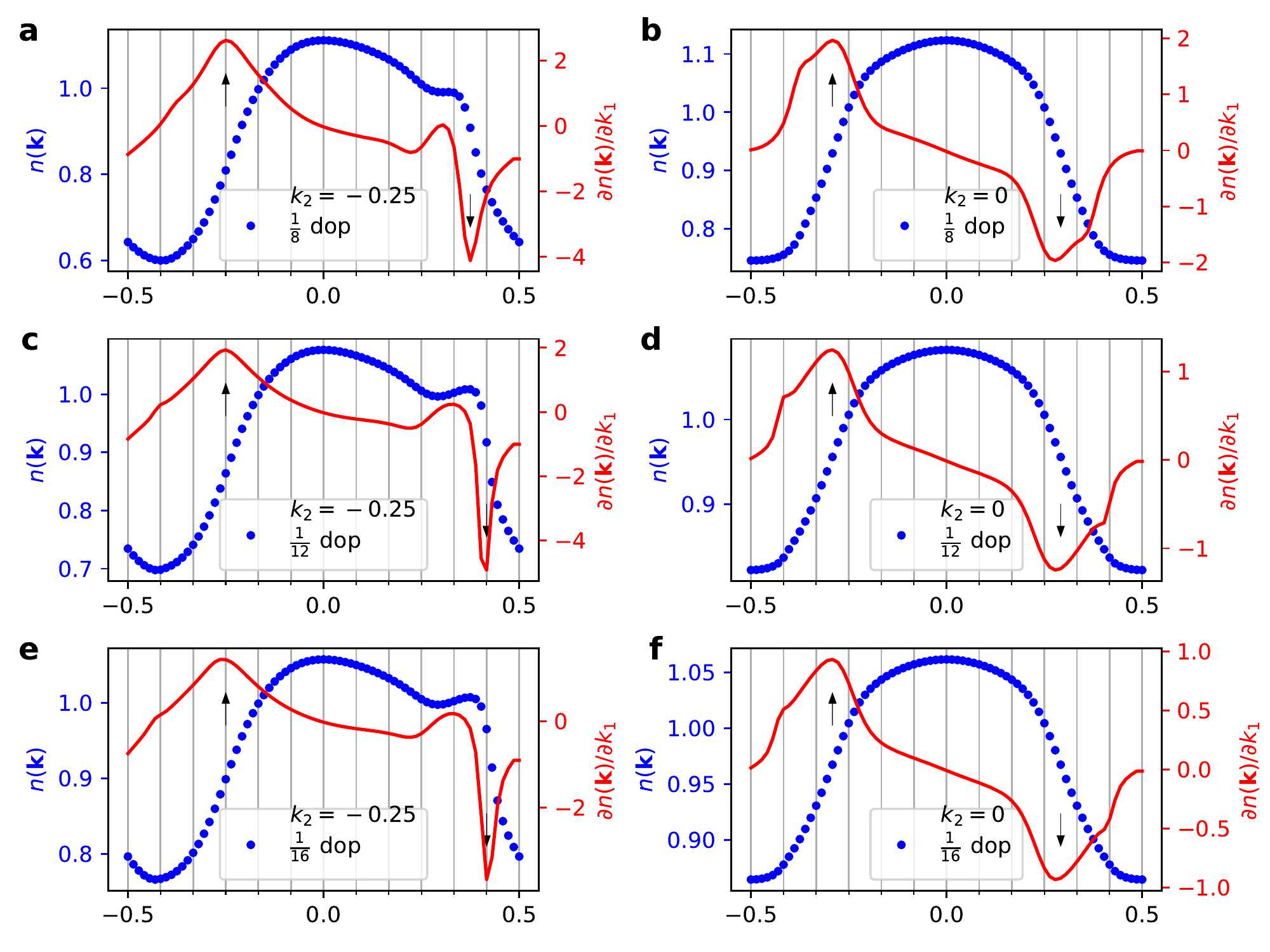}
\caption{Momentum space density $n(\vec{k})$ and its first derivative in the $k_1$ direction $\partial n(\vec{k})/\partial k_1$, with parameters $K/J=0.8$, $t/J=2$ and four-leg ladder length $L_x=72$. In the plot, $k_1$ and $k_2$ are the coordinates of $\vec{k}$ points in the unit of primitive vectors in reciprocal space $\vec{k}=k_1 \vec{b}_1 + k_2 \vec{b}_2$.
As there is inversion symmetry, only two cuts $k_2=-0.25$ and $k_1=0$ are shown.
The peak and dip positions (denoted by black arrows) of $\partial n(\vec{k})/\partial k_1$ give the $\vec{k}_F$s. We have $\vec{k}_F = \pm\frac{7}{24}\vec{b}_1$, $(\frac{3}{8}\vec{b}_1-\frac{1}{4}\vec{b}_2)$ and $(-\frac{1}{4}\vec{b}_1-\frac{1}{4}\vec{b}_2)$ for 1/8 hole doping (figure (a) and (b)), $\vec{k}_F = \pm\frac{7}{24}\vec{b}_1$, $(\frac{5}{12}\vec{b}_1-\frac{1}{4}\vec{b}_2)$ and $(-\frac{1}{4}\vec{b}_1-\frac{1}{4}\vec{b}_2)$ for 1/12 (figure (c) and (d)) and 1/16 hole doping (figure (e) and (f)). The precision of these values are limited by the finite system size.
}
\label{fig4}
\end{figure}

In order to identify the pairing symmetry, we analyze the relative phase of $P_{22}$ and $P_{12}$. It is clear from Fig.~\ref{fig1}(b), that $P_{22}$ and $P_{12}$ have out of phase oscillation. Thus we conclude  that we have found a $d$-wave type PDW.

In addition to the power law decay of the pairing correlation, we also found power law decay of both spin and dimer correlation, corresponding to gapless spin and charge degrees of freedom. In Fig.~\ref{fig3}, we show the log-log plot of those correlations. The spin correlations are defined as $S(i_x-i_{x_{0}})=\sum_{i_y} \left\langle\vec{S}_{(i_{x_{0}},i_y)}\cdot\vec{S}_{(i_x,i_y)}\right\rangle$ and dimer correlations are defined as $D(i_x-i_{x_{0}})=\sum_{i_y} \left( \langle d(i_{x_0},i_y)d(i_x,i_y)\rangle - \langle d(i_{x_0},i_y)\rangle \langle d(i_x,i_y)\rangle \right)$ with $d(i_x,i_y)=\vec{S}_{(i_x,i_y)}\cdot\vec{S}_{(i_{x}+1,i_y)}$ the long direction dimer operator. The spin and dimer correlations show slower power law decay compared with the PDW and also show oscillations, and we also analyze their period by Fourier transformation. We find period $2a_0$ for both spin and dimer at hole doping 1/12 and 1/16, while at doping 1/8, we have period $2a_0$ for dimer but period $4a_0$ for spin. 

We also measured the  Fermi vectors $\vec{k}_F$s, which can be estimated by the singular positions of the density in momentum space $n(\vec{k})$. The momentum space Fermi density is calculated as $n(\vec{k})=\frac{1}{N}\sum_{ij\sigma}e^{i\vec{k}\cdot(\vec{r}_{i}-\vec{r}_{j})}\langle c_{i\sigma}^{\dagger}c_{j\sigma}\rangle$.  Fig.~\ref{fig4} shows $n(\vec{k})$ along different cuts of the Brillouin zone (BZ). We collect all the singular points in $n(\vec{k})$ where the first derivative $\partial n(\vec{k}) / \partial k_1$ has a dip or peak and get the $\vec{k}_F$s shown in the caption of Fig.~\ref{fig4}. Although those estimations of $\vec{k}_F$ are rather crude, we can make a consistent check of the Fermi surface area based on these $\vec{k}_F$s. For example for 1/8 hole doping, we add up the distances between the Fermi crossings along the 3 lines given by $(0, \pm\frac{1}{4})\vec{b}_2$ and multiply by the width of each line which is 1/4 (in the unit of width of the first BZ) to get a total area of $11/24=0.4583 $ (in the unit of area of the first BZ). This is  close to the free fermion value of (1-$p$)/2 = 0.4375 for the 2D Fermi surface. If we focus on the change of Fermi surface area from 1/8 to 1/12 hole doping, we find $1/48$ from this estimate, in precise agreement with what is expected for free Fermions. Thus we conclude that our interacting quasi-1D system retains the Fermi surface structure expected for  Luttinger liquids.

The three crossings of Fermi surface from the measurements of $n(\vec{k})$ correspond to six gapless modes. To verify it, we measured the subsystem entanglement entropy and fit the central charge with formula $S(l,N)=\frac c 6 \log \left( \frac N {\pi} \sin \frac {\pi l} {N} \right) + A$, where $l$ is the length of subsystem, $N$ is the total number of sites, $c$ is the central charge, $A$ is a constant and the $l=1$ entropy gives a very good estimation of it when $N$ is sufficiently large. While we have not reached convergence, we find that central charge is at least 4 and  is  consistent with  central charge of $c=6$ which supports six gapless modes due to three crossings. The details of the estimations are presented in Supplemental Material~\cite{suppl}.

\textit{Discussions}\,---\,
As the period of spin correlation in 1/8 hole doping is two times the period of charge correlation and equals the period of PDW, this is reminiscent of the "antiphase" stripe found in La-based cuprates near 1/8 hole doping, where the onset of pairing correlations coincides with the onset of static spin-stripe order, and they share same periodicity~\cite{berg2009striped}. 
On the other hand, for doping of 1/12 and 1/16, there is no such relation between the various periodicity and the PDW cannot be interpreted as stripes.  While we do not have a clear picture of what controls the PDW period,  we find that the  empirical relation for the wave-vector,  $4\pi p/a_0 $,  works perfectly for $p$=1/8 and 1/12 and within errors for $p$=1/16. This reminds us of the discussion of pairing of electrons on the same side of the Fermi surface to form a  PDW with wave vector $2k_F$ in a one dimensional Luttinger liquid, which was referred to as $\eta$-pairing ~\cite{emery1999}. The power law decay is governed by the exponent $\kappa_\rho+1/\kappa_\rho$ which is always greater than 2 for any value of the Luttinger parameter $\kappa_\rho$~\cite{emery1999}. If we regard our quasi-1D system as a set of 4 interacting 1D Luttinger liquids, the quantity that is fixed is the sum of the $2k_F$ from the Fermi crossing of each band  and it is given by $4[(1-p)/2](2\pi/a_0)$ where the factor 4 accounts for the fractional BZ area taken up by each 1D band.  It is interesting to note that up to umklapp this is just our empirical formula $4\pi p/a_0 $.

We believe that the key reason why we find a PDW upon doping the $J$-$K$ model of the triangular lattice is that we are doping into a spin liquid. While the spin liquid in the undoped system has Fermi surfaces, we do not know whether it is a U(1) spin liquid which has a full Fermi surface, or a Z2 spin liquid 
with a partially gapped Fermi surface. The latter has spinon pairing and it is natural to expect that doping will immediately lead to a pairing state which may be exotic. We indeed find the emergence of an exotic PDW. We do not believe the introduction of the ring exchange term alone is sufficient. We have added the ring exchange term to the $t$-$J$ model on a square lattice and the leading pairing correlator remains uniform d-wave.

In conclusion, we find it encouraging that a dominant PDW correlation emerges upon doping of a model that supports a spinon Fermi surface state. Since this model may be applicable to 1T-TaS$_2$, we continue to encourage experimentalists to dope this material by gating in order not to introduce too much disorder~\cite{he2018spinon}. The existence of fluctuating PDW in a doped Mott insulator model is also encouraging news for the search of fluctuating PDW in underdoped cuprates~\cite{lee2014amperean}.

\begin{acknowledgements}
{\it Acknowledgements}\,---\, 
X.Y. Xu is thankful for the discussion with E.M. Stoudenmire and Donna Sheng.
 The calculations are performed using the ITensor C++ library (version 2.1.1). X. Y. Xu and K. T. L. are thankful for the support of Hong
Kong Research Grant Council through C6026-16W,
16324216 and 16307117.
 K.T.L. is further supported by the Croucher Foundation and the Dr Tai-chin Lo Foundation. P.A.L. acknowledges support by the US Department of Energy, Basic Energy Sciences under grant DE-FG02-03ER46076.
He also thanks the hospitality of the Institute for Advanced Studies at the Hong Kong University of Science and Technology.
The simulation is performed at Tianhe-2 platform at the National Supercomputer Center in Guangzhou.
\end{acknowledgements}

\bibliography{main}
\clearpage
\begin{center}
\textbf{\large Supplemental Material for "Pair Density Wave in the Doped $t$-$J$ Model with Ring Exchange on a Triangular Lattice"}
\end{center}
\setcounter{equation}{0}
\setcounter{figure}{0}
\setcounter{table}{0}
\setcounter{page}{1}
\makeatletter
\renewcommand{\thetable}{S\arabic{table}}
\renewcommand{\theequation}{S\arabic{equation}}
\renewcommand{\thefigure}{S\arabic{figure}}

\setcounter{secnumdepth}{3}

\section{Details of DMRG simulation}
\begin{figure}[h!]
\includegraphics[width=0.9\columnwidth]{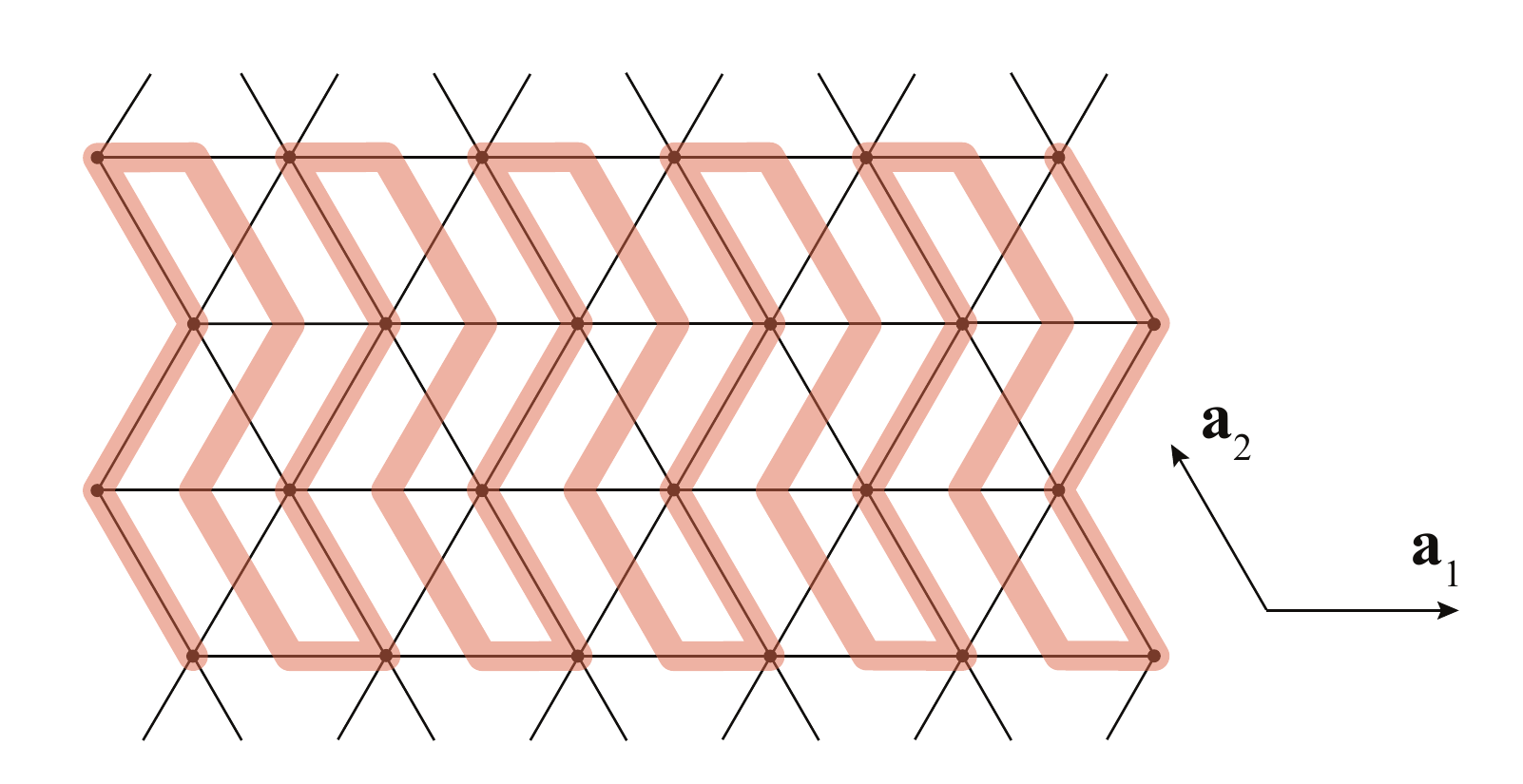}
\caption{DMRG path geometry for a four-leg ladder with length $L_x=6$. Periodic boundary condition is used in the short direction, and open boundary condition is used in the long direction. $\vec{a}_1$ and $\vec{a}_2$ are primitive vectors of the triangular lattice.}
\label{figs1}
\end{figure}

We use DMRG to simulate the $t$-$J$ model with four-spin ring exchange on a triangular lattice as defined in the main text, and we repeat it here for the convenience of discussion.
The total Hamiltonian is
$H= \hat{\mathcal{P}} \left( H_{t-J} + H_K \right) \hat{\mathcal{P}}$, where $\hat{\mathcal{P}}$ excludes doubly occupied states. The $t$-$J$ term $H_{t-J}$ and four-spin ring exchange term $H_K$ are written as
\begin{align}
& H_{t-J} =-t\sum_{\langle i,j\rangle\sigma} \left( c_{i\sigma}^{\dagger}c_{j\sigma}+\text{h.c.} \right) + J\sum_{\langle i,j\rangle}\vec{S}_{i}\cdot\vec{S}_{j}, \\
& H_{K} =K\sum_{\langle i,j,k,l\rangle}\left[\left(\vec{S}_{i}\cdot\vec{S}_{j}\right)\left(\vec{S}_{k}\cdot\vec{S}_{l}\right)+\left(\vec{S}_{j}\cdot\vec{S}_{k}\right)\left(\vec{S}_{i}\cdot\vec{S}_{l}\right)\right. \nonumber \\
 & \hskip 6em \left.-\left(\vec{S}_{i}\cdot\vec{S}_{k}\right)\left(\vec{S}_{j}\cdot\vec{S}_{l}\right)\right],
\end{align}
where $\langle i,j \rangle$ denotes nearest neighbor bond, and $\langle i,j,k,l \rangle$ runs over all compact rhombuses.
The simulation is performed on four-leg ladders on a triangular lattice. As showed in Fig.~\ref{figs1}, periodic boundary condition is used in the short direction, and open boundary condition is used in the long direction. To accelerate the simulation, good quantum numbers are used, and the simulations are performed in the $S^{z}_{\text{tot}}=0$ and  $N_{\text{tot}}=N(1-p)$ subspace. We also exclude the double occupancy states in this subspace. The largest bond dimension we used is 5120, and the largest system size we simulated is with length $L_x=72$, and the corresponding truncation error is less than $10^{-5}$.

\begin{figure}[t]
\includegraphics[width=0.9\columnwidth]{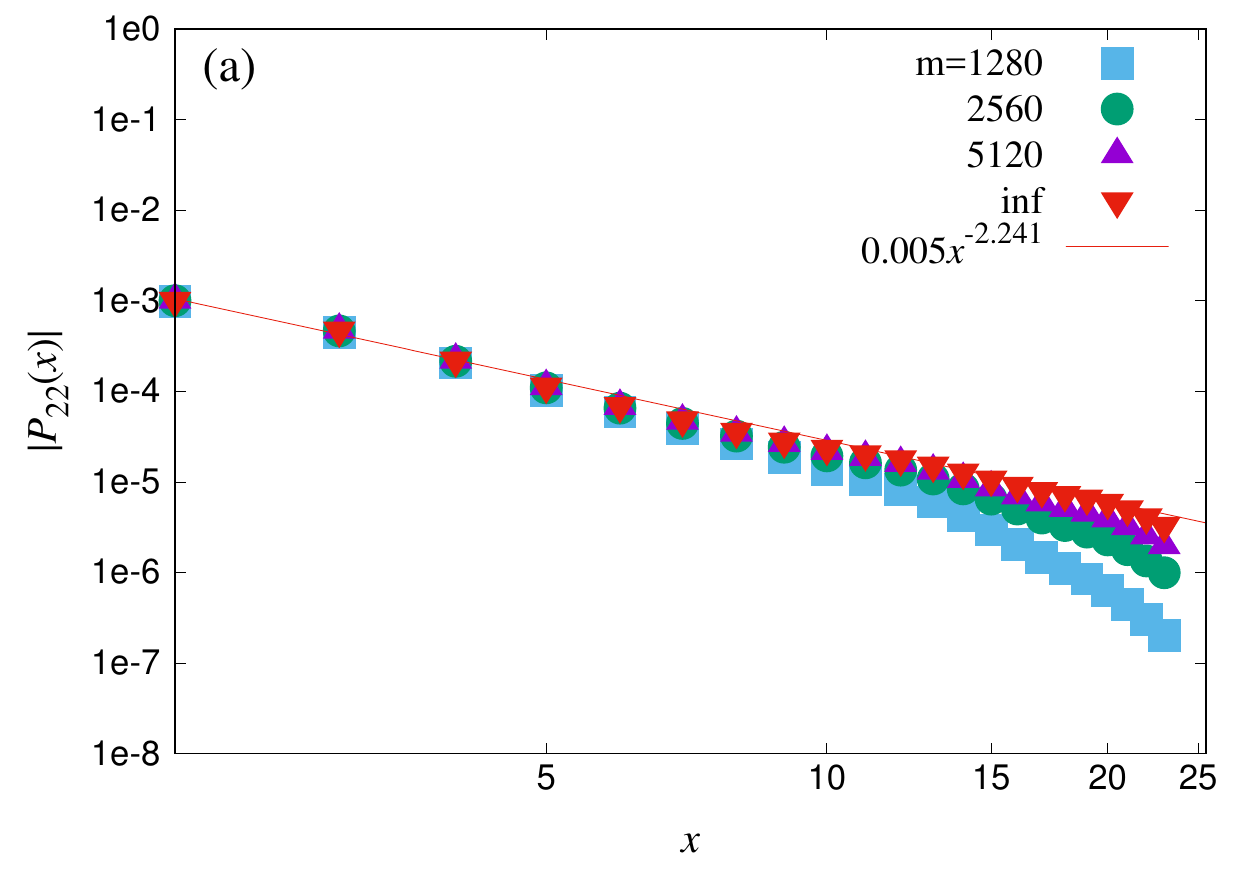}
\includegraphics[width=0.9\columnwidth]{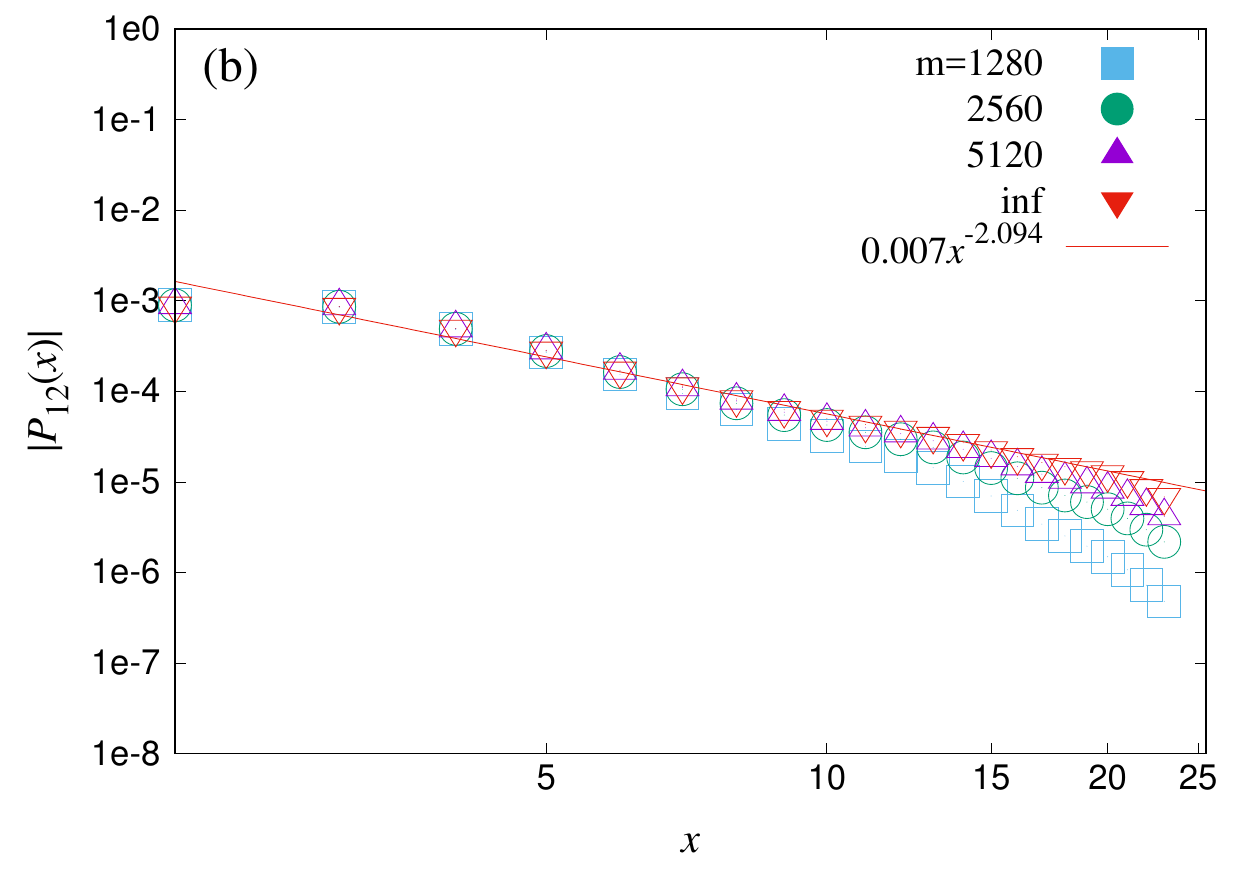}
\caption{Magnitude of pairing correlation for $t$-$J$ model without four-spin ring exchange. Here we have $K/J=0$ and 1/16 doping. (a) is for $P_{22}$ and (b) is for $P_{12}$. The values for bond dimension $m=1280$, $2560$ and $5120$ are showed. The solid symbols denote positive values, while open symbols denote negative values. "inf" denotes values got from the  extrapolation to infinite bond dimension ($\frac 1 m=0$) with a second order polynomial function of $\frac 1 m$.  
The power law fitting (red line) is performed for the extrapolated values (red points).}
\label{figs2}
\end{figure}
\section{Pairing correlations}
We considered both spin singlet and triplet pairing, and found that the spin singlet pairing is always dominant, so we will only focus on the singlet pairing correlation in the following. We define the singlet pairing correlation as 
\begin{equation}
P_{aa'}(i_x-i_{x_{0}})=\sum_{i_y}\left\langle\Delta_{a}^{\dagger}(i_{x_{0}},i_y)\Delta_{a'}(i_x,i_y)\right\rangle,
\end{equation}
where the singlet pairing order parameter $\Delta_{a}(i)$ is defined on nearest neighbor bonds $\Delta_{a}(i)=c_{i,\uparrow} c_{i+\delta_{a}, \downarrow} - c_{i,\downarrow} c_{i+\delta_{a}, \uparrow}$, with $a=1,2$ denoting the long and short direction respectively.  To reduce finite size effects, a summation over the short direction is performed. In the above formula, $(i_x,i_y)$ is the coordinate of site $i$ in the unit of primitive vectors, $i_{x_0}$ is a reference coordinate in the long direction. We take $i_{x_0}>\frac{L_x}{4}$ to reduce boundary effect, and the final results are averaged over several $i_{x_0}$s. We denote the relative distance in the correlator by $x=i_x-i_{x_{0}}$.
\subsection{Pairing correlations of the $t$-$J$ model on a triangular lattice} 
The pairing correlation without four-spin ring exchange is showed here for a comparison. Fig.~\ref{figs2}(a) and Fig.~\ref{figs2}(b) show the pairing correlation $P_{22}$ and $P_{12}$ of a $L_x=48$ system with $K/J=0$, $t/J=2$ and 1/16 hole doping. The data for different bond dimensions are showed here, and the magnitude of the pairing correlations show power law behavior, especially when it is extrapolated to infinite bond dimension. As $P_{22}$ and $P_{12}$ have different signs, they show a $d$-wave type pairing.

\begin{figure}[t]
\includegraphics[width=0.9\columnwidth]{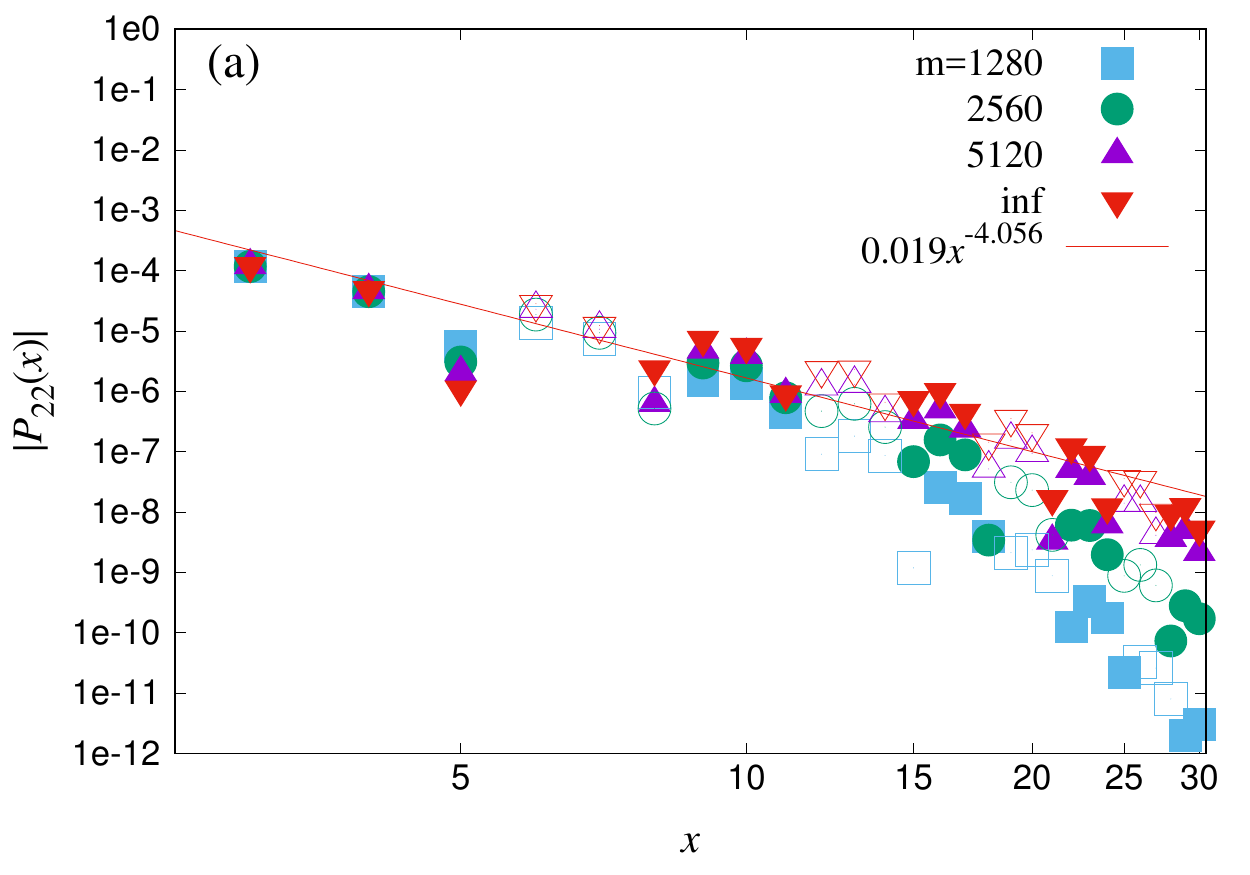}
\includegraphics[width=0.9\columnwidth]{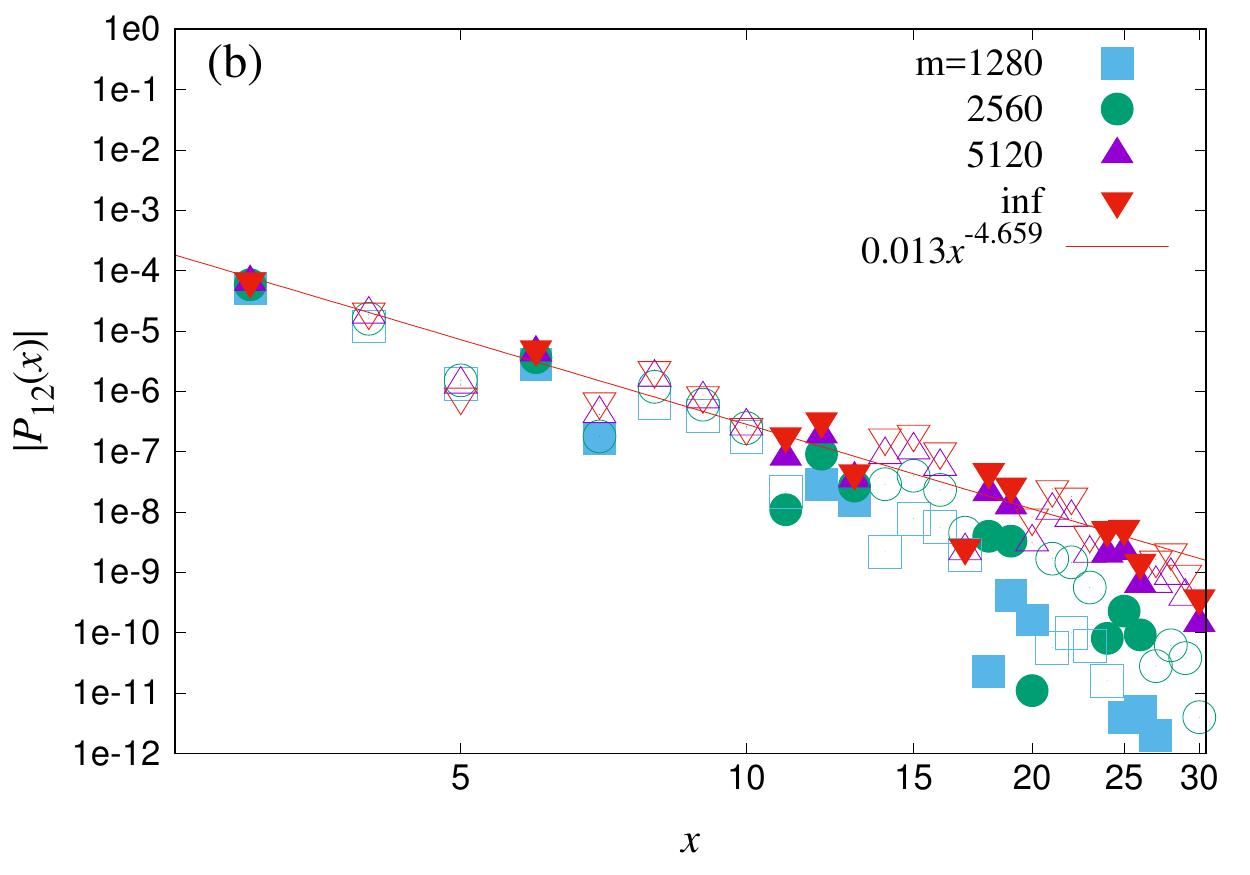}
\caption{Magnitude of pairing correlation for $t$-$J$ model with four-spin ring exchange. Here we have $K/J=0.8$ and 1/16 hole doping. (a) is for $P_{22}$ and (b) is for $P_{12}$. The values for bond dimension $m=1280$, $2560$ and $5120$ are showed. The solid symbols denote positive values, while the open symbols denote negative values. "inf" denotes values obtained from the  extrapolation to infinite bond dimension ($\frac 1 m=0$) with a second order polynomial function of $\frac 1 m$.  
The power law fitting (red line) is performed for the extrapolated values (red points).}
\label{figs3}
\end{figure}
\subsection{Pairing correlations with four-spin ring exchange}
Fig.\ref{figs3}(a) and Fig.\ref{figs3}(b) show the pairing correlation $P_{22}$ and $P_{12}$ of a $L_x=72$ system with $K/J=0.8$, $t/J=2$ and 1/16 hole doping. The data for bond dimension $m=1280$, $2560$ and $5120$ are showed. 
As we see, the pairing correlations with four-spin ring exchange have a significant change.  First, there is sign oscillation in the pairing correlation, and the oscillation of the pairing correlation shows up even at $m=1280$, which indicates the robustness of the PDW. Second, the convergence of the magnitude of the pairing correlation with bond dimension is slow, especially for the long distance part. As we increase the bond dimension, the magnitude of the long distance correlation increases significantly, and it becomes increasingly like power law behavior. Thus we interpret the deviation from power law decay at large distances as being due to a lack of convergence. 

\begin{figure}[b]
\includegraphics[width=1.0\columnwidth]{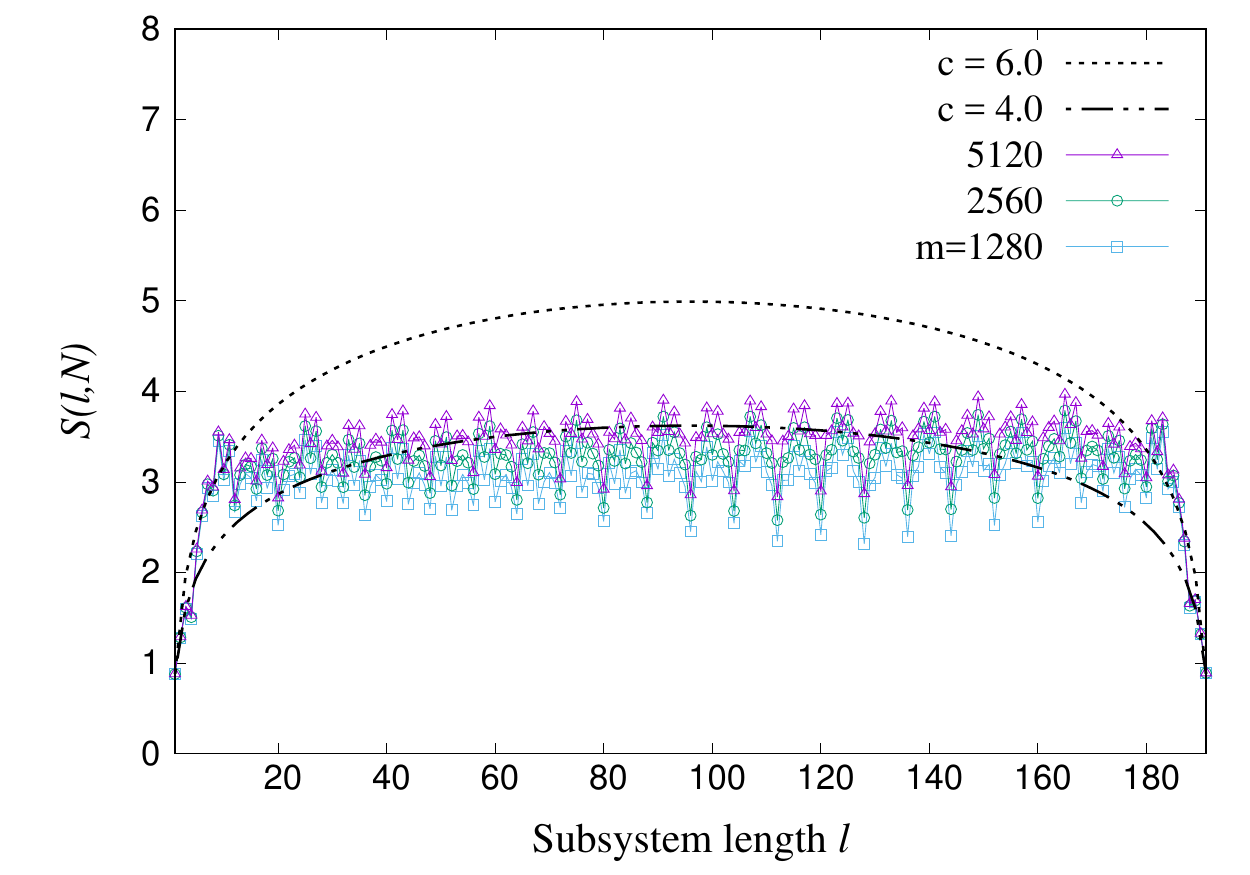}
\caption{Subsystem entanglement entropy of a $L_x=48$ four-leg ladder system with $K/J=0.8$, $t/J=2$ and 1/8 doping. The dashed lines are given by the formula in Eq.~\ref{eq:entropy} with $c=4.0$ and $c=6.0$ with $A$ approximated by $S(l=1,N)$. }
\label{figs4}
\end{figure}

\section{Entanglement entropy}
As we showed in Fig.4 of the main text, there are three crossings of the Fermi surface, and we expect six gapless modes in the system. To analyze the number of gapless modes in the system, we measured the subsystem entanglement entropy, and estimated the central charge by the formula
\begin{equation}
S(l,N)=\frac c 6 \log \left( \frac N {\pi} \sin \frac {\pi l} {N} \right) + A
\label{eq:entropy}
\end{equation}
where $l$ is the length of subsystem, $N$ is the total number of sites, $c$ is the central charge, $A$ is a constant and the $l=1$ entropy gives a very good estimation of its value when $N$ is sufficiently large. 
Fig.~\ref{figs4} shows the subsystem entanglement entropy for a $L_x=48$ system with $K/J=0.8$, $t/J=2$ and 1/8 hole doping. 1/12 and 1/16 hole doping exhibit similar behavior (not shown here). We find that $c$ is clearly larger than 4 and it is consistent with $c=6.0$, especially near the boundary part where the entanglement entropy is well converged. 

\end{document}